\documentclass[letter]{aa}

\usepackage{natbib}
\bibpunct{(}{)}{;}{a}{}{,}
\usepackage{txfonts}

\usepackage{graphicx}

\title{On the abundances of GRO J1655-40\thanks{Based on data obtained at the Very Large Telescope, European Southern Observatory, under program IDs 276.D-5027(A) and 073.D-0473(A).}}

\author{C. Foellmi\inst{1,2} \and T.H. Dall\inst{3} \and E. Depagne\inst{1,4}
}
\institute{European Southern Observatory, 3107 Alonso de Cordova, Casilla 19001, Vitacura, Santiago, Chile \and
Laboratoire d'Astrophysique, Observatoire de Grenoble, BP 53, 38041 Grenoble Cedex 9, France
\and
Gemini Observatory, 670 N. A'ohoku Place, Hilo, HI 96720, USA
\and
Departamento de Astronom\'ia y Astrof\'isica, Pontificia Universidad Cat\'olica de Chile, Campus San Joaqu\'in. Vicu\~na Mackenna 4860 Casilla 306, Santiago 22, Chile.
}

\offprints{C.Foellmi \email{cfoellmi@eso.org}}

\abstract
{The detection of overabundances of $\alpha$-elements and lithium in the secondary star of a black-hole binary provides important insights about the formation of a stellar-mass black-hole. $\alpha$-enhancement might theoretically also be the result of pollution by the nucleosynthesis occurring during an outburst, or through spallation by the jet.}
{We study the abundances, and their possible variations with time, in the secondary star of the runaway black-hole binary GRO~J1655--40, in order to understand their origin.}
{We present a detailed comparison between a Keck spectrum obtained in 1998 found in the literature, archival VLT-UVES data taken in 2004 and new VLT-UVES spectra obtained early 2006. We carefully determine the equivalent widths of different $\alpha$-elements (Mg, O, Ti, S and Si) with their associated uncertainty. We use the well-studied comparison star HD~156098 as well as synthetic spectra to match the spectrum of GRO~J1655--40 in order to determine the abundances of these elements.}
{We see no significant variations of equivalent widths with time. Our fit using HD 156098 reveals that there is significant overabundance of oxygen in all our spectra, but no overabundances of any of the other $\alpha$-elements. Finally, we do not detect the lithium line at 6707~\AA.}
{We show that there is no detected pollution in GRO~J1655--40 after the burst in 2005. Moreover, we argue that uncertainties in the equivalent widths were previously underestimated by a factor of $\sim$3. Consequently, our results challenge the existence of general overabundances of $\alpha$-elements observed in this galactic black-hole binary, and thus the accepted interpretation that they are of supernova origin. The physical cause of the overabundance of oxygen remains unclear.}


\date{Received $<$date$>$, Accepted $<$date$>$}

\titlerunning{On the abundances of GRO J1655-40}

\begin{document}

\maketitle

\section{Introduction}

In 1999, Israelian and coworkers published evidences of overabundances of $\alpha$-elements by a factor 6 to 10 compared to solar, observed in the secondary star of the runaway black-hole \object{GRO~J1655-40} \citep{Israelian-etal-1999}, discovered in 1994 by BATSE \citep{Zhang-etal-1994}. The spectrum has been taken when GRO~J1655-40 was in quiescence, on May 24, 1998 with the 10-m Keck~I telescope. The authors made a LTE modeling of the spectrum (corrected for non LTE effects), and overabundances were explained by a probable pollution by the original supernova, also responsible for the kick velocity of the system. This explanation has been extensively studied, for instance, by \citet{Podsiadlowski-etal-2002} who conclude that the black-hole was formed through a two-steps black-hole formation scenario, with substantial fallback. 

Recently, \citet[][hereafter Paper I]{Foellmi-etal-2006b} have presented an analysis of archival VLT-UVES spectra of GRO~J1655-40 obtained in 2004, when the system was in quiescence. Paper I is dedicated to the problem of the distance of GRO~J1655-40, but we found during the analysis of the spectrum that overabundances of $\alpha$-elements (in particular Mg) were not detected, casting doubt on the overabundances in the Keck spectrum. If the disappearance of such overabundances between 1998 and 2004 was real, it was certainly associated also with a short timescale phenomenon. A mechanism of flare-pollution is known to occur in Cataclysmic Variables \citep{Stehle-Ritter-1999}, but seems to have never been observed in microquasars. Models invoke accretion disk nucleosynthesis \citep[e.g.][]{Mukhpadhyay-Chakrabarti-2000} that might produce in particular large amounts of lithium \citep[e.g.][]{Martin-etal-1994,Guessoum-Kazanas-1999}, or misalignment of the jet with the accretion disk spin \citep[e.g.][]{Butt-etal-2003}.

GRO~J1655--40 is an ideal laboratory for studying these questions. Before the Keck observation in 1998, the source has been in outburst the year of its discovery (1994, see reference above), in 1995 and 1996 \citep[see ][for a chronology of the activity of GRO~J1655-40 from 1994 to 1997]{Orosz-etal-1997}. Interestingly, GRO~J1655-40 has been in quiescence between 1999 and early 2005 where a new outburst occurred \citep[see e.g.][]{Brocksopp-etal-2006}. It has reached again its complete quiescent level in October 2005 (ATel \#644\footnote{See {\tt http://www.astronomerstelegram.org/}}). Moreover, GRO~J1655-40 is considered in the literature as a misaligned microquasar \citep[e.g.][]{Maccarone-2002} since the inclination angle of the system is around 70$^{\circ}$ \citep{Greene-etal-2001}, while the jet angle is at 85$^{\circ}$ according to \citet{Hjellming-Rupen-1995}. {\it However,} in Paper I we have shown that the distance of GRO~J1655-40 is much smaller than previously used values. Thus, adopting a distance of 1.0~kpc, as we inferred in Paper I, the radio data of \citet{Hjellming-Rupen-1995} imply a jet angle of 72$^{\circ}\pm 2$. In this case, GRO~J1655-40 is not misaligned.

We have conducted a new VLT-UVES program at the beginning of 2006. We present in this Letter the comparison between the results of \citet{Israelian-etal-1999}, and the UVES spectra taken in 2004 and 2006, to check the possible presence of lithium, and to examine whether overabundances of $\alpha$-elements in the secondary star of GRO~J1655-40 exist, and whether they are variable. 

\section{Observations}

VLT-UVES spectra taken in 2004 are described in Paper I. We obtained new spectra of GRO~J1655-40 with VLT-UVES on the nights Feb. 18, March 7, 8 and 19, 2006. In total, five spectra with a central wavelength of 5800\AA\ were acquired, covering the exact same spectral domain as the 2004 spectra: from 4785 to 5755\AA\ and 5835 to 6805\AA. One spectrum was also obtained with a central wavelength of 8600\AA, covering the regions 6680 to 8540\AA\ and 8650 to 10080\AA, matching the wavelength range presented by \citet{Israelian-etal-1999}. The resolving power of the spectra is 45~000. Each individual spectrum has a S/N of about 40 to 60 in the continuum. They have been reduced the same way as described in Paper I, and shifted to the heliocentric restframe. In order to achieve a higher Signal-to-Noise ratio, we have cross-correlated, shifted and combined the spectra together. The resulting spectra have a S/N between 100 and 150 in the continuum.

\section{Analysis}

As confirmed in Paper I, the secondary star in GRO~J1655-40 is a F6IV star with plenty of absorption lines, and a rotational velocity of 94 km~s$^{-1}$. As an immediate consequence it is extremely hard to find line-free regions in the spectrum, and thus to safely determine the continuum level.

\subsection{Temporal variations of equivalent widths}

In order to carefully measure the equivalent widths (EWs), we have: (1) Fit the continuum of the whole  spectrum with a low-order polynomial choosing small line-free regions of the spectrum. (2) Cross-correlate and shift the 2006 spectra to match the 2004 velocity. (3) Linearly rebin the 2006 spectra to the exact same wavelength dispersion of the 2004 spectra. (4) Define {\it common} wavelength limits to the line wings to measure the EWs. The EWs are summarized in Table~\ref{table_ew}, where the values of \citet{Israelian-etal-1999} are also reproduced. We have computed the internal uncertainties of the EWs using the recent work by \citet{Vollmann-Eversberg-2006}. We also tested our EW measurements with gaussian convolution and binning before the continuum fit. It usually gives smaller values of EWs, since pre-continuum fit operations tends to smear out the lines with the continuum, thus lowering the contrast with the lines.

Our tests show clearly that the continuum level is the main source of uncertainty, rather than the quality of the spectra themselves. Given the S/N ratio being around 100 or slightly above, we find impossible to define the continuum level to better than 1\%, especially given the large number of absorption lines, and the rotational broadening often causing blends. This is a crucial point, since {\it none} of the lines in Table~\ref{table_ew} are actually isolated lines. We have thus recomputed the EWs with identical wavelength limits but artificial vertical shifts of $\pm$0.01 continuum units, and square added the uncertainties to the previous ones. This contribution to the uncertainty largely dominates, by 60 to 95\%. 

Moreover, it appeared meaningless to measure EWs smaller or close to 100 m\AA, even if we rebin the spectra to a resolution of 5~000 to increase S/N ratio. For instance, the blend of iron lines at $\sim$6633\AA\ can have an EW between 0 and 200~m\AA\ depending on the order of the fitting polynomial. Thus, our uncertainties are systematically larger by a factor of $\sim$3 compared to that of \citet{Israelian-etal-1999}, although our spectra are of better quality (S/N $\sim$100--150 vs 35) and a slightly higher resolving power (45 000 vs $\sim$30 000, assuming a 2.5 pix resolution element for the Keck spectrum\footnote{Some of the overabundances reported in \citet{Israelian-etal-1999} were obtained with an even lower-resolution spectrum by \citet{Shahbaz-etal-1999}}). 

\begin{table}
\centering
\caption{Equivalent widths (EWs, in units of m\AA) of various absorption lines observed in 1999, 2004 and 2006. The values of 1999 are taken unchanged from \citet{Israelian-etal-1999}. A $^{+}$ sign indicates values actually quoted by the authors from the work by \citet{Shahbaz-etal-1999}. Horizontal lines delimitate groups of lines for which only one EW can be measured.}
\begin{tabular}{cl|ccc}\hline \hline
Line & Ion & 1999 & 2004 & 2006 \\ 
     &     & (m\AA) & (m\AA) & (m\AA) \\ \hline
5167.32 & Mg I & -  & 940 $\pm$ 70 & 890 $\pm$ 65 \\ \hline
5172.68 & Mg I & -  & 420 $\pm$ 50 & 475 $\pm$ 45 \\ \hline
5183.60 & Mg I & -  & 510 $\pm$ 50 & 505 $\pm$ 50 \\ \hline
6633.42 & Fe I &                   &              &              \\
6633.75 & Fe I & 105 $\pm$ 15      & $\lesssim$100 & $\sim$100 \\
6634.12 & Fe I &                   &              &              \\ \hline
6663.24 & Fe I & 80 $\pm$ 15 $^{+}$& $\sim$100 & $\sim$100 \\
6663.44 & Fe I &                   &              &              \\ \hline
6677.99 & Fe I &130 $\pm$ 15 $^{+}$& $\gtrsim$100 & $\lesssim$100 \\ \hline
6749.03 & Cr I & 80 $\pm$ 15 $^{+}$& $\sim $100 & 180$\pm$60 \\ 
6750.16 & Fe I &                   &              &              \\ \hline
7771.95 & O I  &               & &               \\
7774.17 & O I  & 1050 $\pm$ 80 & & 1190 $\pm$ 100 \\
7775.39 & O I  &               & &               \\ \hline
7780.56 & Fe I & 90 $\pm$ 15   & &  160 $\pm$ 30 \\ \hline
8446.24 & O I  &               & &               \\
8446.35 & O I  & 530 $\pm$ 50  & &  460 $\pm$ 60 \\
8446.75 & O I  &               & &               \\ \hline
8467.15 & Ti I &               & &               \\
8468.40 & Fe I &               & &  560 $\pm$ 110 \\
8468.50 & Ti I &               & &               \\ \hline
8693.93 & S I  & 230 $\pm$ 25  & &  $\sim$100 \\
8694.62 & S I  &               & &               \\ \hline
8728.01 & Si I & 250 $\pm$ 25  & &  $\gtrsim$100 \\
8728.59 & Si I &               & &               \\ \hline
8736.02 & Mg I & 200 $\pm$ 20  & &  150 $\pm$ 60 \\
8736.03 & Mg I & & & \\ \hline
\end{tabular}
\label{table_ew}
\end{table}

Given the uncertainties, it can be seen that there is no obvious EW variations between our UVES spectra in 2004 and 2006. The comparison between 1998 and 2004 shows no systematic significant variations neither. Thus, the overall metallicity and the $\alpha$-element abundances of GRO~J1655-40 have remained constant since 1998.  Moreover, we do not detect the lithium $\lambda$6707 line at a level above our uncertainties in any of our spectra.

\subsection{$\alpha$-element abundances}

In order to study the abundances of the secondary star in GRO~J1655-40, we performed both a LTE spectral synthesis of selected regions using MOOG \citep{Sneden-1973}, and spectral comparison with artificially broadened stellar templates, thereby bypassing problems with NLTE effects which affect in particular some oxygen lines. For the templates we used the UVES POP database of spectra \citep{uvespop}\footnote{{\tt http://www.sc.eso.org/santiago/uvespop/}}, taking advantage of the fact that this provides spectra taken with the same instrument and setup. As explained in Paper I, and as confirmed by \citet{Israelian-etal-1999}, any of the template stars explored in Table~\ref{table_ew} of Paper I can be used without introducing noticeable uncertainties in the derived abundances. However, for consistency with Paper I, we use exclusively the slowly rotating star \object{HD 156098} as a template \citep[F6IV, V=5.537 mag, T$_\mathrm{eff}$=6480 $K$, $\log g$=3.94, Fe/H=0.09;][]{Edvardsson-etal-1993}, which corresponds almost exactly to the model used by \citet{Israelian-etal-1999}. \citet{Bensby-etal-2005} confirmed the metallicity of HD~156098 and found a slight enhancement of the $\alpha$-elements ([$\alpha$/Fe]$ = 0.10$; average of Mg, Si, Ca, Ti --- [O/H]$\sim$0.0--0.1). The synthesis and spectral subtraction was done using STARMOD \citep{Barden-1985,Montes-etal-1995b,Montes-etal-2000}. Given the large $\alpha$-enhancement of GRO~J1655-40 ([$\alpha$/Fe]$ \sim 1.0$) found by \citet{Israelian-etal-1999} we would expect to see clear enhancements of lines of the $\alpha$-elements with respect to the template spectrum.  

\begin{figure}
\centering
\includegraphics[width=0.9\linewidth]{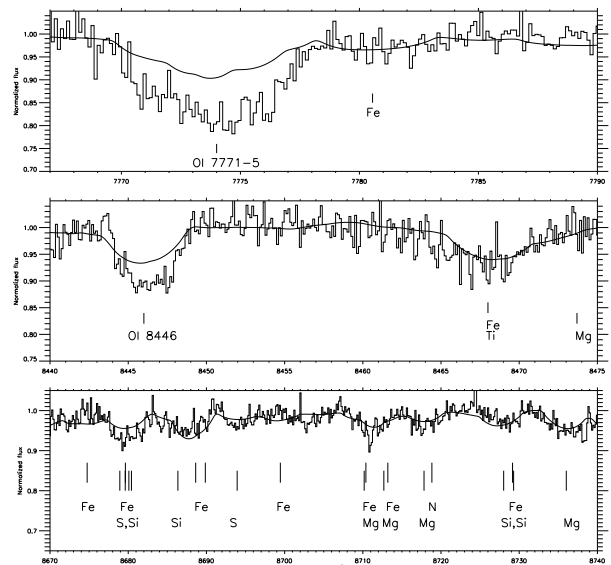}
\caption{UVES 2006 spectrum of the oxygen triplet $\lambda$\,7771-5 (top), the \ion{O}{i} $\lambda$\,8446 line (middle), and the region around 8700~\AA, containing many lines of the $\alpha$-elements. We have indicated the major constituents of the line blends, plotting the $\alpha$-element indicators and identifications below those of the other elements. The overabundance of oxygen is obvious, while the abundances of the other $\alpha$-elements are comparable with the abundances in HD~156098 (solid line). These three regions correspond almost exactly to the plots of the 1998 Keck-HIRES spectrum \citep[][Fig.~1]{Israelian-etal-1999}, with a few modifications due to UVES inter-order gaps.}
\label{oxygen}
\end{figure}

In Fig.~\ref{oxygen} we show the general good match of comparing the spectra of GRO~J1655-40 with the broadened template spectrum, using the same spectral regions as in \citet{Israelian-etal-1999}.  We confirm the large overabundance of oxygen, which is obvious from both sets of oxygen lines. However, we do not see any evidence for general $\alpha$-enhancement: the region around 8700~\AA\ seem to match the template very well, meaning that the abundances of Ca, Mg, Si, Ti, and S are at most only slightly overabundant, corresponding to the [$\alpha$/Fe]=0.10 found in HD~156098.  

\begin{figure}
\centering
\includegraphics[width=0.9\linewidth]{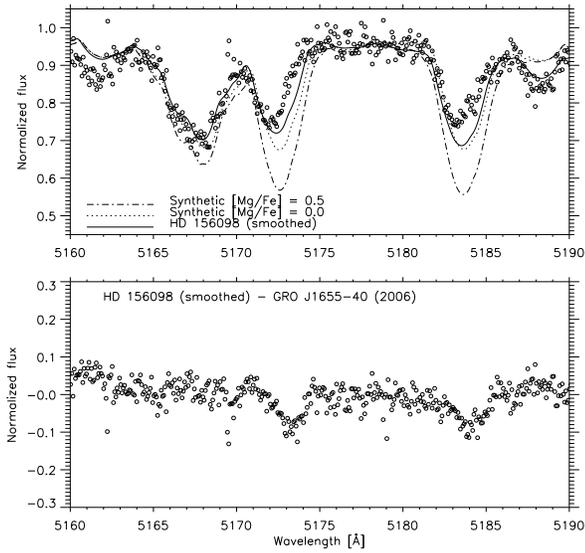}
\caption{One of the single UVES 2006 spectrum of GRO~J1655-40 (symbols) with the F6IV broadened stellar template of HD~156098 (solid line). Also plotted are two model spectra with [Mg/Fe]=0.0 (dotted line) and [Mg/Fe]=0.5 (dot-dashed line). The apparent wavelength shift is caused by the blending of several lines with the Magnesium ones. It has however no influence on Mg abundances. Below are plotted the residuals after subtracting the HD~156098 template spectrum from the GRO~J1655-40 spectrum.}
\label{match}
\end{figure}

To check this result, we investigated two other spectral regions around the \ion{Mg}{i} and \ion{S}{i} triplets. For the \ion{Mg}{i} triplet $\lambda\lambda$\,5167,5172,5183 we also calculated synthetic spectra for a range of magnesium abundances with the same parameters as those of HD~156098. The result is shown in Fig.~\ref{match}, which again reveals no obvious $\alpha$-enhancement, as the lines match the HD~156098 template very well. The \ion{Mg}{i} triplet is known to exhibit NLTE effects, so one must be cautious when interpreting the synthetic spectra. Nevertheless, the [Mg/Fe]=0.0 model seem to fit HD~156098 quite well, while the [Mg/Fe]=+0.5 model is clearly a bad fit to either star. Finally, the relative strength of the three lines can be used as a rough indicator of abundance, with the blue-most line being weaker than the other two for overabundances larger than [Mg/Fe]$\sim$~0.3. Changing $T_\mathrm{eff}$ and $\log g$ of the model has little effect on the abundances.

\begin{figure}
\centering
\includegraphics[width=0.9\linewidth]{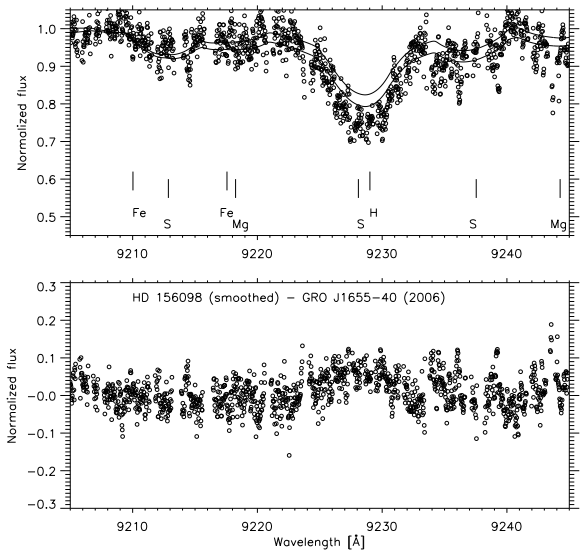}
\caption{UVES 2006 spectrum of GRO~J1655-40 (symbols) with the F6IV broadened stellar template of HD~156098  (lower continuous line) in the region of the \ion{S}{i} triplet. The upper continuous line show the same stellar template, but on which the difference between two acceptable normalizations of the GRO~J1655-40 spectrum has been applied. It shows the amplitude of uncertainties in the continuum level in this region, and that in both cases HD~156098 is a providing a good match. Below are plotted the residuals after subtracting the HD~156098 spectrum from the best-fitting GRO~J1655-40 spectrum.}
\label{s1}
\end{figure}

The far red region of the strong \ion{S}{i} $\lambda\lambda$\,9212,9228,9237 triplet coincides with a broad-winged hydrogen absorption line. When normalizing the spectrum of GRO~J1655-40 we naturally had to exclude the area contaminated by the hydrogen line, which introduces some ambiguity in the solution. Using two equally satisfactory continuum normalizations, it is clear from Fig.~\ref{s1} that sulphur is not overly abundant.

\section{Discussion and conclusions}

We are led to conclude that the $\alpha$-element overabundances found by \citet{Israelian-etal-1999} were overestimated, except for oxygen, which we confirm is clearly overabundant. The physical cause of this overabundance in the secondary star of GRO~J1655-40 remains unclear.

\begin{acknowledgements}
We thank the ESO Director General Dr.~C.~Cesarsky for accepting our DDT run early 2006. THD thanks Garik Israelian and Rafael Rebolo for useful discussions and hospitality. We thank I.F. Mirabel for useful advices and support, and the anonymous referee for useful comments.
\end{acknowledgements}

\bibliographystyle{aa}

\end{document}